# An Adaptive and Multi-Service Routing Protocol for Wireless Sensor Networks

Jaydip Sen
Innovation Lab, Tata Consultancy Services Ltd.
Bengal Intelligent Park, salt Lake Electronic Complex, Sector V
Kolkata – 700091, INDIA
Jaydip.Sen@tcs.com

***Abstract*—Wireless Sensor Networks (WSNs) are highly distributed networks consisting of a large number of tiny, low-cost, light-weight wireless nodes deployed to monitor an environment or a system. Each node in a WSN consists of three subsystems: the sensor subsystem which senses the environment, the processing subsystem which performs local computations on the sensed data, and the communication subsystem which is responsible for message exchange with neighboring sensor nodes. While an individual sensor node has limited sensing region, processing power, and energy, networking a large number of sensor nodes give rise to a robust, reliable, and accurate sensor network covering a wide region. Thus, routing in WSNs is a very important issue. This paper presents a query-based routing protocol for a WSN that provides different levels of Quality of Service (QoS): energy-efficiency, reliability, low latency and fault-tolerance-under different application scenarios. The algorithm has low computational complexity but can dynamically guarantee different QoS support depending on the requirement of the applications. The novelty of the proposed algorithm is its ability to provide multiple QoS support without reconfiguration and redeployment of the sensor nodes. The algorithm is implemented in network simulator ns-2 and its performance has been evaluated. The results show that the algorithm is more efficient than some of the currently existing routing algorithms for WSNs.**

***Keywords-wireless sensor networks, routing, quality of service, energy-efficiency, reliability, latency.***

## I. INTRODUCTION

Recent convergence of technological and application trends has resulted in an exceptional level of interest in wireless ad-hoc networks and in particular in wireless sensor networks (WSNs). The push was provided by rapid progress in computation and communication technology as well as the emerging field of low cost, reliable, MEMS-based sensors. The pull was provided by numerous applications that can be summarized under the umbrella of computational worlds, where the physical world can be observed and influenced through the Internet and WSN infrastructures. Consequently, there have been a number of vigorous research and development efforts at all levels of development and usage of WSNs, including applications, operating systems, architectures, middleware, integrated circuits, and systems.

Typically, WSNs contain hundreds or thousands of sensor nodes that have the ability to communicate with each other and also with an external base station (BS) [1]. While the sensor nodes have limited sensing region, processing power and energy, networking a large number of such nodes gives rise to a robust, reliable and accurate sensor network covering a wider region. Since the sensor nodes are energy-constrained, a typical deployment of a WSN poses many challenges and necessitates energy-awareness at all layers of the networking protocol stack. For example, at the network layer, it is highly desirable to find methods for energy-efficient route discovery and relaying of data from the sensor nodes to the BS such that the lifetime of the network is maximized.

Design of an efficient routing protocol is a particularly challenging task due to some of the characteristic features of such networks. First, due to large number of sensor nodes, a global addressing scheme cannot be applied and hence IP-based protocols are not applicable. Second, in contrast to other wireless networks, applications in WSNs require flow of sensed data from multiple sensor nodes (sources) to a particular BS (sink). Third, sensor nodes are tightly constrained in terms of energy, processing, and storage capacities and a routing protocol should minimize the resource consumption. Finally, data collected by many sensors in WSNs is typically based on common phenomena. Hence, there is a high probability that these data have some redundancy. Such redundancy needs to be exploited by the routing protocols so as to improve the energy-efficiency and bandwidth utilization.

A large number of routing protocols for WSNs currently exists in the literature. These protocols can be classified into multipath-based, query-based, negotiation-based, and quality-of-service (QoS)-based depending on the application of the protocol [2]. In this paper, a query-based adaptive routing protocol is proposed that can satisfy multiple QoS requirements such as reliability and latency. Since the algorithm does not involve any complex computation, it is highly energy-efficient.

The rest of the paper is organized as follows. Section II discusses some of the currently existing routing protocols for WSNs. Section III presents the details of the proposed algorithm. Section IV describes the simulation scenarios and the results obtained. Finally, Section V concludes the paper while highlighting some future scope of work.

## II. Related Work

As mentioned in Section I, a large number of routing protocols have been proposed in the literature. This Section discusses some of these schemes.

Intanagonwiwat et al. have proposed a popular data aggregation protocol for WSNs [3]. It is based on the concept of directed diffusion and is particularly useful in scenarios, where the sensor nodes themselves generate the queries for data sensed by other nodes, instead of all queries originating from the BS. The performance of directed diffusion is affected by a number of factors, e.g., the position and the number of source nodes, the topology of the WSN etc. While it can lead to reduction in energy consumption, it cannot be applied in a scenario where continuous monitoring of an event is required.

In situations where the amount of data to be exchanged is small and the quality of the paths is not important, an alternative technique is rumor routing [4]. In rumor routing, the source information and sink interest trajectories are random straight rays emanating from their respective origin. An agent emanates out of each node and propagates the data or interest. As the agents move, the corresponding data is stored in all nodes that the agents pass through. Although, rumor routing avoids flooding, it performs well only when the number of events is small.

Heinzelman et al. have proposed a family of adaptive protocols called sensor protocols for information via negotiation (SPIN) that disseminate information at each node to every other node in the network [5]. The algorithm uses negotiation and resource adaptation to address the deficiencies of the flooding approach. However, the data advertisement mechanism in this algorithm cannot guarantee delivery of data.

Ye et al. have proposed a minimum cost forwarding algorithm (MCFA) for setting up paths to a sink in a WSN [6]. Each node maintains the least cost estimate from itself to the BS, and broadcasts each message to its neighbors. This process is repeated till the BS is reached. Although MCFA is an efficient protocol, it invokes an expensive backoff algorithm in the setup phase in order to avoid multiple and frequent updates received at the nodes which are far away from the BS.

Geographic hash table (GHT) is a system based on data centric storage and is suitable for large scale WSNs [7]. It hashes keys into geographic coordinates and stores a (key, value) pair at the sensor node nearest to the hash value. The computed hash value is mapped onto a unique node consistently so that queries for the data can be routed to the correct node. The data is distributed among the nodes such that it is scalable and the storage load is balanced.

Burri et al. have proposed a data gathering protocol for WSNs that caters to the requirements of periodic data collection with ultra-low power consumption [8]. The protocol presents a coordinated approach to MAC-layer design, topology control, and efficient routing to reduce energy wastage in communication, in which using a tree-based network structure, packets are reliably routed towards the sink.

Cugola et al. have presented a context-aware and content-based routing (CCBR) protocol that is explicitly designed for multi-sink, mobile WSNs [9]. It adopts a probabilistic receiver-based approach to routing and uses content-based addressing to effectively support data-centric communication. It also takes into account the contexts of the sensor nodes to filter data.

Pandana et al. have proposed a class of energy-aware routing algorithm that takes into account the connectivity of the nodes in WSNs to mitigate the problem of network partitioning due to failure of nodes [10]. The protocol utilizes the concept of node importance, which characterizes the algebraic connectivity of the remaining graph when a node failure occurs. Several properties of the proposed routing algorithm have been presented and the effectiveness of the protocol has been demonstrated by extensive simulations.

Gnawali et al. have presented two principles for wireless routing: data path validation and adaptive beaconing [11]. In the data path validation phase, data traffic quickly discovers and fixes routing inconsistencies. The adaptive beaconing phase extends the trickle algorithm [12] for routing control traffic by reducing route repair latency and sending fewer beacons.

Awad et al. have proposed the virtual cord protocol (VCP) for efficient routing in WSNs [13]. It involves greedy routing on the cord and does not require exact location information of the nodes. The protocol is scalable since the nodes only require information about their direct neighbors. However, the protocol is not fault-tolerant.

In addition to the above work, power-efficient gathering in sensor information systems (PEGASIS) [14], and adaptive threshold-sensitive energy efficient protocols (APTEEN) [15] are two very well-known WSN routing algorithms. In the following section, the details of the proposed routing algorithm are presented.

## III. Proposed Routing Algorithm

This section presents the details of the proposed routing algorithm. In contrast to most of the current algorithms, the proposed algorithm does not attempt to optimize on a single parameter e.g., energy, latency and reliability. Instead, it attempts to integrate multiple QoS parameters within a single framework. The adaptive feature of the protocol makes it suitable for a numerous application scenarios without requiring any reconfiguration of the sensor nodes. Before we discuss the details of the algorithm, we present a set of assumptions which is required to be valid for the algorithm to work.

### A. *Assumptions*

For the design and working of the algorithm the following assumptions are made. Most of the assumptions are practical and can be easily realized in real-world WSN deployments. The assumptions are as follows:

(i) A node in the WSN, known as the *sink* node, floods the network with query messages periodically. In response to the query from the sink node, some nodes send back their responses. These nodes are called the *source* nodes. (ii) The wireless links connecting the sensor nodes are assumed to be symmetric. In other words, if a sensor node $x$ is within the communication range of sensor node $y$, then $y$ is also within the communication range of $x$. (iii) A query-based data gathering model is used. The source nodes send responses to the sink node only on receiving the query. (iv) No in-network processing is assumed. However, the proposed protocol will work even is situations where in-network processing and data aggregation is done at some intermediate sensor nodes. (v) The sink has an external power source. The other sensor nodes are battery-driven. (v) All the nodes are assumed to be static and each node has a unique identifier. (vi) The sensor nodes are assumed to be capable of transmitting at any one of the two pre-defined power levels -- one for the short- range communications and other for the long-range communications.

The proposed routing protocol is designed to cater to applications belonging to four different QoS classes: (i) *normal applications* with no guarantee (i.e. reliability) of packet delivery and latency constraints, (ii) *reliable applications* which require guaranteed delivery of query messages from the sink to the sources and response messages from the sources to the sink, without any constraints on latency of packet delivery, (iii) *delay-sensitive, real-time* applications with *no specific reliability requirements*, which are time-critical, and (iv) *delay-sensitive, real time* applications with *specific reliability requirements*, for which packets from the sink to the sources and in the reverse path must traverse within a specific delay bounds and with a very high probability of successful delivery at the destination.

### B. Algorithms for different application classes

In this subsection, the details of the proposed algorithm for the four different classes of applications are presented.

*Case1: Applications with no relaibility and latency constraints*

This module of the algorithm is applicable for applications which are neither delay-sensitive nor have any reliability requirement. When the sink node requires information from the source nodes, it broadcasts a DATA_REQ message, which is essentially an IP packet. For normal applications, the *type of service* (ToS) bit in the DATA_REQ header is set to "00" so that all nodes receiving this packet understand that the packet is for a *normal* application. A typical DATA_REQ packet header contains the following additional information: (i) energy level of the node, (ii) minimum hop-count of the node from the sink, (iii) node identifiers of three neighbors of the node that have least-hop counts from the sink. As the routing information is updated in all the nodes, each node creates and populates a table maintained locally. This table is called *forwarding information table* (FIT). The FIT of a node $x$ contains the following information: (a) node identifier of each of its neighbor nodes from which it has received DATA_REQ packet, (b) the energy level of each neighbor, (c) for each neighbor of node $x$, node identifiers of three neighbors that have least-hop counts from the sink. This information is updated based on the DATA_REQ packets arriving at the nodes. In addition to the records for each of its neighbors, every node also keeps its own information in its FIT. At the network bootstrapping time, all hop-counts are set an arbitrarily large value. Figure 1 shows the structure of an FIT.

Let the hop-count field in the DATA_REQ header from a neighbor node $k$ of node $i$ be $L_k$, and the hop-count of node $i$ from the sink be $H_i$. When a DATA_REQ packet traverses from say, node $k$ to node $i$, FIT of node $i$ is updated as follows:

(i) If $L_k + 1 < H_i$, the value of $H_i$ is substituted by that of $L_k$. If FIT of node $i$ currently has no record for node $k$, a new entry for node $k$ is created. The $H_i$ field is updated based on the DATA_REQ packet, and the packet is broadcasted in the neighborhood of node $i$.

(ii) If $L_k + 1 = H_k$, a new entry for node $k$ is created in FIT of node $i$. The hop-count field is updated with the $H_i$ value in DATA_REQ packet and the packet is broadcasted in the neighborhood of node $i$.

(iii) If $L_k + 1 > H_i$, the DATA_REQ packet is dropped at node $i$ without any further broadcasting, because the distance of node $k$ from which the DATA_REQ has arrived is further away from the sink than node $i$.

When node $i$ receives a DATA_REQ packet for which it is the intended source node, it sends the response in the form of a DATA_REP packet. For routing the DATA_REP packet to the sink, node $i$ uses the following algorithm.

| Neighbour ID (N) | Energy Level (E) | Hop-Count to Sink (H) | Forwarder 1 ID | Forwarder 2 ID | Forwarder 3 ID |
|---|---|---|---|---|---|
| | | | | | |

Figure 1. A typical forwarding information table (FIT)

*Routing Algorithm 1*

Let $N_i$ be the set of neighbors of a node $j$ where $i$ ε {1, 2, 3 …., $n$}. Let $F_{im}$ be the set of forwarders for node $i$ in the FIT where $m$ ε {1, 2, 3}. Let $E_i$ represents the energy level of node $i$.

*Step 1*: Select $N_x$, $N_y$, $N_z$ ε $N_i$ from the FIT of node $i$ such that their corresponding hop-counts $H_x$, $H_y$, and $H_z$ are of the minimum value.

*Step2*: Select $N_k$, where $k$ ε {$x$, $y$, $z$} and $E_k$ is maximum.

if $F_{km} \cap N_i == \Phi$, then forward the data packet to $N_k$.

else $N_i = N_i - N_k$. Repeat *Step 1.*

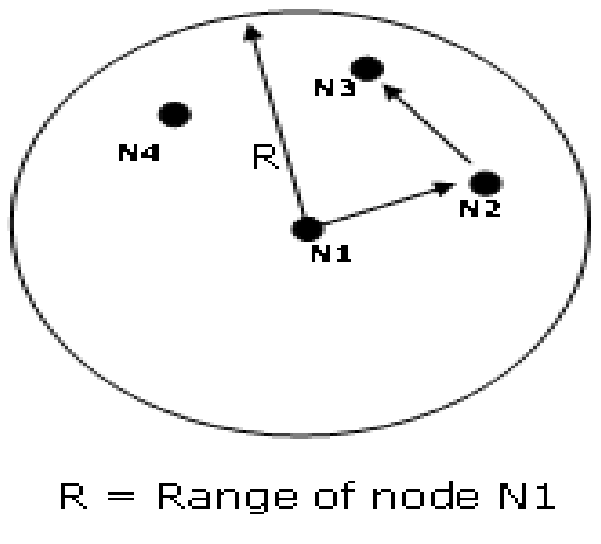

Figure 2. Illustration of the roles of forwarding nodes

To understand the role of the forwarder nodes, let us refer to Figure 2. Nodes $N_2$, $N_3$, and $N_4$ are the forwarder nodes for node $N_1$. In other words, DATA_REQ sent by the sink node reaches $N_1$ and it is further forwarded to nodes $N_2$, $N_3$ and $N_4$. Let $N_2$ and $N_3$ are two-hops and one-hop away from the sink. When $N_1$ sends the DATA_REP message to the sink, it chooses $N_2$, $N_3$ and $N_4$ as forwarders based on their hop-count records in FIT. If $N_2$ has higher energy than $N_3$ and $N_4$, then $N_1$ forwards the DATA_REP packet to $N_2$. When the DATA_REP packet reaches $N_2$, it will not be forwarded to $N_3$, based on the hop-count information in $N_2$'s FIT. This reduces delay and prevents unnecessary energy-drain in the sensor nodes since $N_1$ could have directly sent the packet to $N_3$.

The routing algorithm works based on information available in a neighborhood and does not require any network-wide global information. Moreover, route discovery can be done dynamically and does not involve maintenance of any route-cache. This makes the algorithm computationally simple with a minimal memory overhead.

*Case 2: Reliable applications without any delay constraints*

This module of the algorithm is applied for applications which have high reliability requirements for packet delivery. However, delay is not an important issue for these applications. Reliability is achieved by enhancing the (i) reliability of the routing paths and (ii) reliability of packet delivery.

Using single path routing, path reliability is ensured while minimizing the energy consumption in the sensor nodes. However, to achieve higher reliability in packet delivery, multipath routing strategy is employed. Each node employs the *link layer ack* mechanism to detect any failed nodes in its neighbourhood. If an acknowledgment does not reach within a pre-defined time, the neighbour node is assumed to have failed.

There algorithm works in three steps: (i) primary path selection, and (ii) selection of alternate paths at the source, and (iii) alternate path selection at the intermediate nodes.

1. *Primary Path Selection*

The source node *i* checks FIT to find its upstream neighbor (the path from source to sink is the upstream path; the reverse path is the downstream path) with least hop-count from the sink, and forwards the packet to that neighbour if $E_i \geq E_{threshold}$, where $E_i$ is the energy of node *i* and $E_{threshold}$ is the minimum energy-level required for a node to receive and forward the data packet. If the energy-level of a neighbor node falls below $E_{threshold}$, node *i* deletes the corresponding entry for that node from its FIT, and chooses the next upstream neighbor in its FIT. In case of a node failure, the intermediate downstream node finds a new path to the sink by selecting the node with the least hop-count from the source after checking the entries in its FIT.

2. *Alternate paths selection at the source node*

The source node selects its upstream neighbors as follows. Select $N_{1a}$, $N_{2a}$ ε ($N$ - $N_p$) where $N$ is the set of all upstream neighbors of node *i*, $N_p$ is the upstream neighbors in the primary path and $N_{ia}$ and $N_{2a}$ are the upstream neighbors for the source node in the first alternate path and second alternate path respectively. $N_{1a}$ and $N_{2a}$ are also chosen on the basis of their least hop-count information maintained in the FIT of node *i*.

3. *Alternate paths selection at the intermediate nodes*

Since wireless communication is inherently broadcast in nature, the header information of a packet forwarded by a node is accessible to all the nodes in its neighborhood. Looking at the header, it is possible for a node to know whether it lies on the path from a particular source to the sink. Based on these information collected over a period of time, the intermediate nodes construct *path construction tables* (PCTs). In addition to the primary path, PCTs are also utilized in identifying alternate paths from source to the sink. When an intermediate node receives a packet, the node consults its FIT to select a node and compares it with the PCT to ensure that the selected node is not in the path for that particular source node. A typical format of a PCT is depicted in Figure 3.

| Node_ID ($C_i$) | Source_Addr ($C_{isa}$) | Destn_Addr ($C_{ida}$) |
|---|---|---|
| | | |

Figure 3. A typical path construction table (PCT)

After the primary path, first and second alternate paths are created at the source and the intermediate nodes, the following routing algorithm is used for sending packets from the source nodes to the sink node.

*Routing Algorithm 2*

Let $N_i$ be the set of neighbors of a node *j* where *i* ε{1, 2, 3….*n*}. Let $C_i$ be the list of records of nodes in the PCT which are in the path of a particular source-destination pair. Let $P_{sa}$ and $P_{da}$ be the source and destination address in the DATA_REP packet from the source to the destination node.

Step 1: Select $N_x$ ε $N_i$ from the FIT such that its corresponding hop-count $H_x$ is the minimum.

Step 2: If ($C_i \cap N_x == \Phi$)
choose $N_x$ as the forwarding node;
make the corresponding updates in the PCT;
else if ($C_{isa} == P_{sa}$ && $C_{ida} == P_{da}$)
$N_i = N_i - N_x$ ;
repeat Step 1;
else
choose $N_x$ as the nest forwarding node;
make the corresponding entry in the PCT;

The above routing algorithm does not require complex computation as simple comparison of the tables at the intermediate nodes is required to find the next-hop node for forwarding. This requires less energy since use of beacon signals is not required to find the next-hop nodes.

*Case3: Delay-sensitive applications with no reliability*

This module of the algorithm is for applications which have stringent latency bounds, but do not have any specific lower bound on the reliability of packet delivery. To minimize the delay, the nodes transmit at higher power-level so that the number of hops is reduced. The algorithm is as follows:

*Routing Algorithm 3*

Let $N_i$ be the set of neighbors of a node *j* where *i* ε {1, 2, 3 …. , *n*}. Let $T_i$ represents the average waiting time for a packet in the queue for the node *i*. The information about the buffer occupancy (queue length) of each neighbor is broadcasted in node *i*'s neighborhood when FIT information are exchanged. An additional field in the FIT table is created for this purpose.

The queue length in the buffer of a node serves as an estimator of the expected waiting time in that node. The algorithm has the following steps:

*Step 1*: Select $N_x$, $N_y$, $N_z$ ε $N_i$ from the FIT of node $i$ such that their corresponding hop-counts $H_x$, $H_y$, $H_z$ are minimum.

*Step 2*: Select $N_k$, $k$ ε $\{x, y, z\}$ such that $T_k$ is minimum.

Since the largest component of the overall delay that a packet experiences is caused due to waiting in the buffers at the intermediate nodes, the delay is minimized by choosing those forwarder nodes that have least waiting time (or smaller queues of packets in buffer).

*Case 4: Delay-sensitive applications with reliability*

Finally, this case is applicable for applications which are not only time-critical but also require high reliability of packet delivery. The routing module for this case works in the following three steps.

*1. Primary path selection*

Let $N_i$ be the set of neighbors of a node $j$ where $i$ ε $\{1, 2, 3 …., n\}$. Let $T_i$ represents the average waiting time for a packet in the queue for the node $i$. Select $N_x$, $N_y$, $N_z$ ε $N_m$ from the FIT of node $i$ such that their corresponding hop-counts $H_x$, $H_y$, $H_z$ are minimum.

*2. Selection of alternate path at the source*

The source selects the upstream neighbors from its FIT as follows: Select $N_a$ ϵ $(N - N_p)$ where $N_p$ is the upstream neighbor of the source node in the primary path and $N_a$ is the upstream neighbor for the source in the alternate path having the next least average waiting time.

*3. Alternate path selection at intermediate nodes*

The method of selection of forwarding nodes for generation of alternate path at the intermediate nodes is the same as in Case 2 described earlier. However, unlike in Case 2, the node selection is based on the average waiting time of the packets in the buffers of the intermediate nodes and not on the hop-count.

## IV. SIMULATIONS AND RESULTS

The proposed protocol is implemented on network simulator *NS*-2 version 2.3.2 [16]. To study and analyze the performance of the protocol, two metrics are chosen: (i) *average dissipated energy* in sensor nodes, (ii) *average latency* in message communication. Average dissipated energy is the ratio of total energy dissipated in the network to the number of packets received by the sink over a given period of time. The second metric measures the average one-way delay a packet experiences while traversing from a source to the sink. The performance of the proposed protocol is studied with respect to the above two metrics with varying network sizes. Each simulation is carried out with 10 different topologies and the final results are presented based on the average of the result sets. The number of nodes in the network is increased from 50 to 150 with 25 nodes being added in every step of increment. Initially, 50 sensors nodes are randomly distributed on a 70m * 70 m area. The area of deployment is increased as more sensor nodes are added so that the node density is maintained constant. The radio range for each sensor is kept at 15m for simulation of the first two cases (Cases 1 and 2) of the protocol. The transmission power of the nodes is kept at low in these two cases since there is no latency constraint. In the other two cases (Cases 3 and 4), a longer radio range of 30 m is chosen for each node to satisfy the delay constraints. In each simulation run, one node is chosen as the sink and three nodes are randomly chosen as the sources. To ensure that not all the reply packets from the sources are lost, each source sends three reply packets corresponding to every query packet it receives.

Figure 4 shows the average dissipation for different network sizes. Case 1 is the most energy-efficient. Case 2 consumes more power than Case 1 as it uses multiple paths for routing. Case 3 requires higher transmission range to reduce transmission delay and hence consumes more power than Case 2. Case 4 leads to maximum dissipation of power as it ensures highest level of QoS both in terms of reliability and latency.

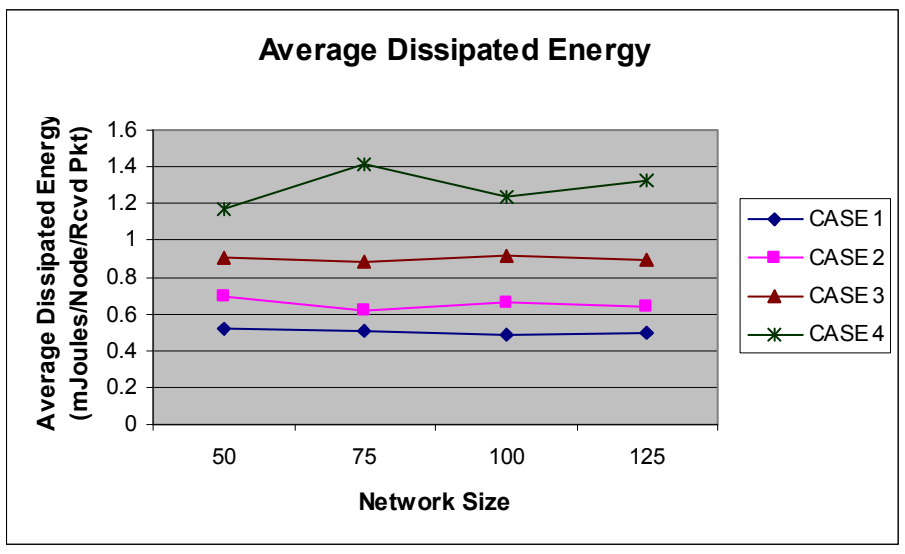

Figure 4. Average dissipated energy in a node with varying network sizes

Figure 5 shows the average delay for the four cases. Case 3 results in the minimum latency since it does not guarantee any reliability. Case 2 has the highest latency since it identifies multiple paths for reliable packet delivery. Although Case 4 uses higher transmission power, it has more latency than Case 3 because of multipath identification at the source. Case 1 causes more delay than Case 4 as it uses less transmission power. However, since construction of multiple paths is not required, it has less latency compared to Case 2.

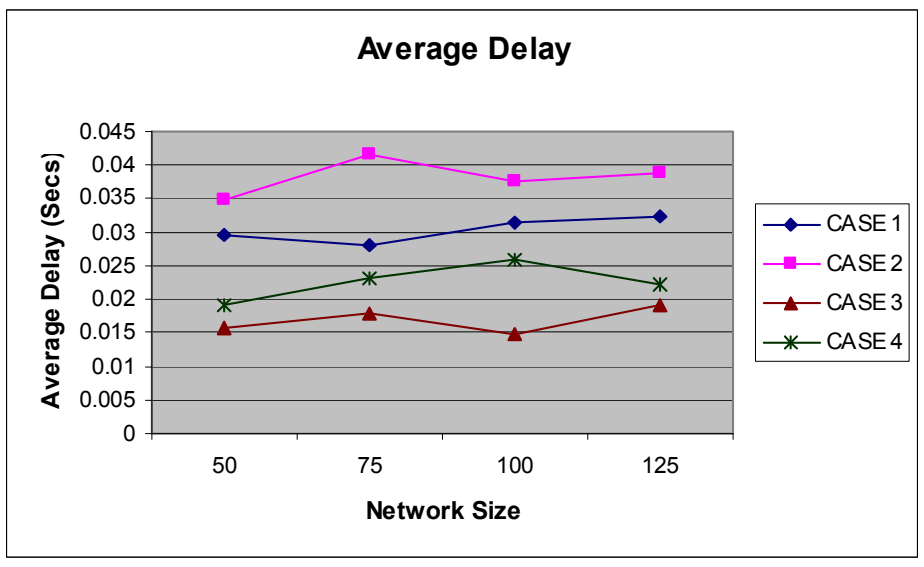

Figure 5. Averge delay experienced by a message in different network sizes

To study the performance of the proposed protocol from reliability perspective, the protocol is simulated in scenarios where sensor nodes fail after the DAT_REQ packets are broadcasted. In this case, the number of reply packets reaching the sink node is taken as the measure of reliability of the protocol. Case 2 always guarantees 100% delivery of the DATA_REP packet to the sink since any failed node is detected beforehand and the packets are routed through nodes in an alternate path. Since Case 1 and Case 3 do not use alternate paths some packets are lost. Case 4 uses multiple paths but some packets are lost due to the delay constraints. To observe the effect of failed nodes, simulations are carried out with 50, 75, 100 and 125 nodes with 10% and 20% of failed nodes in the network. Experiments are repeated 10 times and the average results are presented in Figure 6 and Figure 7. It may be observed that Case 4 gives the highest reliability among the three cases because of the use of alternate paths in routing.

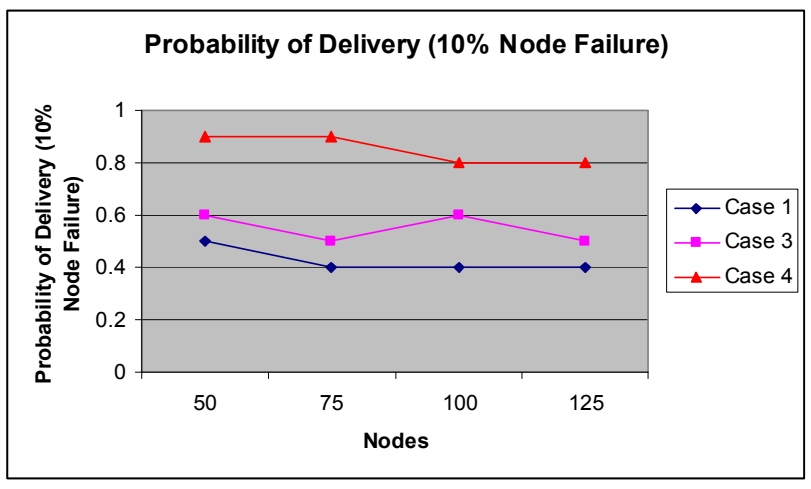

Figure 6. Probaility of successful packet delivery with 10% of nodes failed

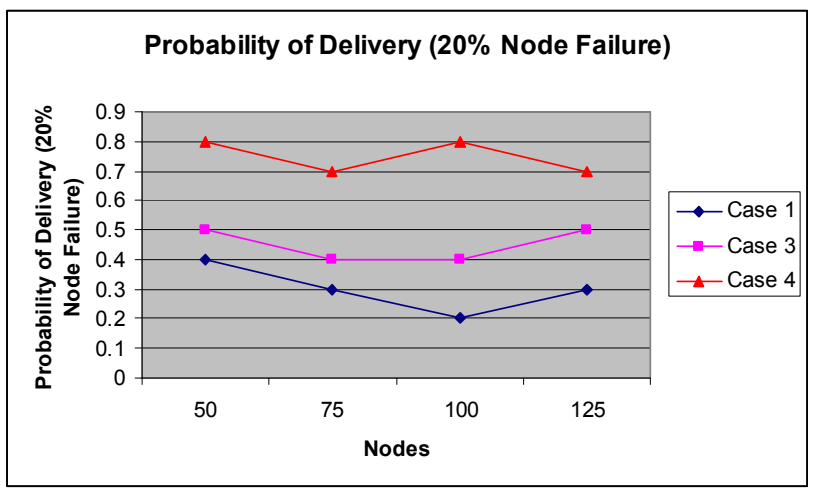

Figure 7. Probaility of successful packet delivery with 20% of nodes failed

It may be observed from Figures 6 and 7 that Case 3 provides more reliability in terms of packet delivery to the sink in comparison to Case 1. Case 3 involves less number of nodes for forwarding a packet from a source to the sink. Thus with the same percentage of failed nodes, the probability that Case 1 will encounter a failed node in its routing path will be higher than that in Case 3, resulting in less packet delivery in Case 1.

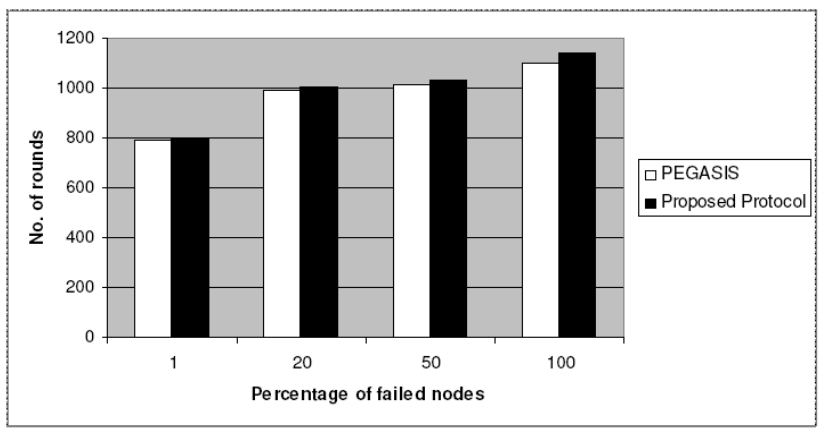

Figure 8. Comparison of the proposed algorithm with PEGASIS

Finally, the proposed algorithm (Case 4) is compared with the PEGASIS [14] protocol, where all sensor nodes can access the BS directly. Random graphs are generated with 100 nodes distributed over a 50m*50m region. The BS is placed at (25, 150) with each node having a range of 110 so that transmission from each node can directly reach BS. Figure 8 shows the comparisons with varying percentage of failed nodes. It is clear that the proposed algorithm provides larger network life-time.

## V. CONCLUSION

In this paper, a query-based, adaptive routing protocol is presented that satisfies QoS requirements such as reliability and latency. For ensuring path reliability the algorithm uses multiple paths from source nodes to the sink nodes and for guaranteeing data reliability it sends multiple copies of the same message. The latency is minimized by enabling the sensor nodes to transmit with more power. The algorithm is simulated on network simulator *ns-2* and its performance is compared with an existing protocol. As a future scope of work, we plan to make the query-driven protocol an event-driven one.

## REFERENCES

[1] I. F. Akyildiz, W. Su, Y. Sankarasubramaniam, and E. Cayirci, “A survey on sensor networks”, *IEEE Communications Magazine*, Vol. 40, No. 8, pp. 102-114, August 2002.

[2] J.N. Al-Karaki and A.E. Kamal, “Routing techniques in wireless sensor networks: a survey”, *IEEE Wireless Communication Magazine*, Vol. 11, No. 6, pp. 6-28, December 2002.

[3] C. Intanagonwiwat, R. Govindan, and D. Estrin, “Directed diffusion: a scalable and robust communication paradigm for sensor networks”, in *Proc. of ACM MOBICOM 2000*, pp. 56-67, August 2000.

[4] D. Braginsky and D. Estrin, “Rumor routing algorithms for sensor networks”, in *Proc. of ACM Workshop on Wireless Sensor Networks and Applications*, pp. 22-31, Septmber 2002.

[5] W.R. Heinzelman, J. Kulik, and H. Balakrishnan, “Adaptive protocols for information dissemination in wireless sensor networks,” in *Proc. of ACM MOBICOM*, pp. 174-185, August 1999.

[6] F. Ye, A. Chen, S. Lu, and L. Zhang, “A scalable solution to minimum cost forwarding in large sensor networks”, in *Proc. of IEEE ICCCN 2001*, pp. 304-309, October 2001.

[7] S. Ratnasamy, B. Karp, S. Shenker, D. Estrin, R. Govindan, L. Yin, and F. Yu, “GHT: a geographic hash table for data-centric storage”, *ACM Journal on Mobile Networks and Applications*, Vol. 8, No. 4, pp. 427-442, August 2003.

[8] N. Burri, P. VonRickenbach, and R. Wattenhofer, “Dozer: ultra-low power data gathering in sensor networks”, in *Proc. of the 6th Int. Conf. on Information Processing in Sensor Networks,* Cambridge, Massachusetts, USA, pp. 450-459, 2007.

[9] G. Cugola and M. Migliavacca, “A context and content-based routing protocol for mobile sensor networks”, in *Proc. of the 6th European Conf. on Wireless Sensor Networks*, Cork, Ireland, pp. 69-85, 2009.

[10] C. Pandana and K.J. Ray Liu, “Robust connectivity-aware energy-efficient routing for wireless sensor networks”, *IEEE Trans on Wireless Communications*, Vol. 7, No. 10, October 2008, pp. 3904 – 3916.

[11] O. Gnawali, R. Fonseca, K. Jamieson, D. Moss, and P. Levis, “Collection tree protocol”, in *Proc. of the 7th ACM Conf. on Embedded Networked Sensor Systems*, Berkeley, CA, USA, pp. 1-14, 2009.

[12] P. Levis, N. Patel, D. Culler, and S. Shenker, “Trickle: a self-regulating algorithm for code maintenance and propagation in wireless sensor networks”, in *Proc. of the USENIX NSDI,* California, USA, March 2004.

[13] A. Awad, C. Sommer, R. German, and F. Dressler, “Virtual cord protocol: a flexible DHT-like routing service for sensor networks”, in *Proc. of the 5th IEEE Int. Conf. on Mobile Ad-hoc and Sensor Systems (IEEE MASS 2008)*, Atlanta, USA, pp. 133-142, September 2008.

[14] S. Lindsey and C.S. Raghavendra, “PEGASIS: power-efficient gathering in sensor information systems”, in *Proc. of IEEE Aerospace Conference*, Vol 3, pp. 1125-1130, 2002.

[15] A. Maneshwar and D.P. Agrawal, “APTEEN: a hybrid protocol for efficient routing and comprehensive information retrieval in wireless sensor networks”, in *Proc. of International Parallel and Distributed Processing Symposium (IPDPS)*, pp. 195-202, April 2002.

[16] Network Simulator NS-2. URL: http://www.isi.edu/nsnam/ns.